**Broad band noise and stochastic resonance in the CDW conductors $K_{0.30}MoO_3$ and $K_{0.30}Mo_{(1-x)}W_xO_3$ (x=0.001 and x=0.002)**

J. Dumas, J. Marcus

Institut Néel, CNRS et Université Joseph Fourier, BP 166, F-38042 Grenoble Cedex 9, France

**Abstract**

We have investigated nonlinear characteristics and broad band noise in the quasi-one dimensional charge-density-wave conductors $K_{0.30}MoO_3$ and $K_{0.30}Mo_{(1-x)}W_xO_3$ (x=0.001 and x=0.002) When the amplitude of a bipolar rectangular voltage pulse excitation is modulated by adding a white noise perturbation, the average pulse amplitude is smaller than the unperturbed one and shows a non monotonous behaviour when the pulse amplitude is increased. Our findings suggest that coupling between internal broad band noise and external white noise may lead to an increase of phase coherence of the charge-density-wave.

**Introduction**

The depinning of a charge density wave (CDW) in quasi-one dimensional conductors has been extensively studied in the past decades [1]. At low electric field, the CDW is prevented from moving by pinning to randomly distributed impurities. Above a depinning threshold $E_t$, the CDW condensate starts to slide. The sliding of the CDW results in nonlinear conductivity accompanied by generation of a large low frequency broad band noise (BBN) with spectral density $S(f) \sim f^{-\alpha}$ ($\alpha \sim 1$) and high frequency coherent oscillations, the so-called narrow-band noise (NBN) which coexists with the BBN. The NBN is characterized by sharp peaks in the noise spectrum. BBN has been studied mainly in $TaS_3$ [2, 3] and $NbSe_3$ [4, 5]. In $K_{0.30}MoO_3$, the BBN is Gaussian, increases above $E_t$ then saturates for large dc bias [6, 7].

Various models have been proposed for the origin of BBN. In the simplest picture, presented by Bhattacharya et al. [8, 9], and Maeda et al. [10], the BBN originates from fluctuations of the impurity pinning force due to deformations of the sliding CDW. The local threshold $E_t$ over a phase coherent region fluctuates in time. Slightly different values of the threshold voltages lead to fluctuations in the chordal resistance R=V/I. In other models, the BBN comes from configurational rearrangements involving flow of CDW defects [11].

Internal degrees of freedom of the CDW manifest in frequency-locking experiments in combined dc+ac excitations [12]. Locking of the internal frequency or synchronization to the external frequency of an ac field $V=V_{dc}+V_{ac}\sin\omega t$ leads to interference phenomena in sliding CDW's characterized by plateaus in the dc V(I) characteristics. In the regime of mode-locking, a reduction of the BBN has been observed [13]. Remarkable memory effects associated with the CDW conduction have been observed, namely pulse sign memory and pulse duration memory effects: As first shown by Gill [14], when an applied current pulse has the same polarity as the preceding pulse, the voltage response is fast. If the polarity of the current pulse has an opposite polarity from the preceding pulse, during the previous depinning, the voltage response is sluggish.

When a series of identical square voltage pulses are applied, the phase of the oscillation at the end of each pulse is found to be independent of the pulse length. The CDW seems to remember or learn the duration of the preceding pulse, as first reported by Fleming and Schneemeyer [15]. More recently, it has been shown by numerical simulations that phase organization memory can be improved by adding a physically plausible type of stochastic noise to the system [16].

The interplay between internal and external frequencies has been studied in detail [12]. This is not the case for the interplay between BBN and an external noise. In order to get some insight into unexplored stochastic aspects of the CDW dynamics and into the depinning process we have studied the CDW voltage response V(t) to bipolar symmetrical rectangular current pulses when a large white noise perturbation is added in the widely studied pure and W-doped quasi-one dimensional conductors $K_{0.30}MoO_3$. The pulse amplitude is modulated by noise with a modulation depth ranging from 20% up to 80 %. When the white noise perturbation is added on the transient voltage pulse excitation, the average pulse amplitude response is found to be smaller than the unperturbed one and shows a non monotoneous behaviour when the pulse amplitude is increased. Our results seem to indicate that coupling between BBN and external white noise leads to an increase of phase coherence of the charge-density-wave.

**Experimental results**

The experimental results were obtained on single crystals of pure and W-doped $K_{0.30}MoO_3$ platelets with typical dimensions of 2x1x0.2 mm$^3$ parallel to the cleavage plane and elongated along the high conductivity *b*-axis. V(I) characteristics and conduction noise measurements were performed at 77K with the standard four-probe configuration. Electrical contacts were made by evaporating silver pads on a freshly cleaved surface. Platinum wires were attached by silver paste onto the electrode contacts. In all cases, the current contacts covered the ends of the sample. Contact resistance was 2 Ohm at 300K. For all samples, the distance between voltage contacts was 1mm. The threshold field for depinning the CDW in pure $K_{0.30}MoO_3$ is sample dependent [17]. The lowest value reported was 15 mV/cm at 77 K. In the present investigations, we have studied pure samples with threshold field values ranging from 95.8 to 1000 mV/cm. For W-doped samples, the isoelectronic element W is substituted on Mo sites and acts as a weak pinning center. The CDW coherence length becomes very short at low tungsten concentrations [18].

The arrangement consists of Tektronix TDS 5034 B Digital Phosphor Oscilloscope and Agilent 33120 A function generator. Bipolar rectangular periodic voltage pulses were applied to the sample. Large resistance was applied in series with the pulse generator to achieve a current driven arrangement. The rectangular shape of the voltage can be written: $V(t) = A$ for $0<t<T/2$ and $V(t) = -A$ for $T/2 <t<T$ with a period T=2ms. Since at 77K approximately 1ms is necessary to reach a steady state voltage [19], the experiments were performed with a pulse width $\tau$=1 ms.

The pulse amplitude is modulated by adding a Gaussian white noise (bandwidth 10 MHz) with a modulation depth ranging from 20% up to 80 %. The modulation depth is defined as the ratio of the peak-to-peak value of the modulation signal amplitude to the pulse amplitude V=2A. In this case, the noise amplitude increases linearly with the excitation V(t). The pulse response was averaged over 2048 scans.

Shown in Figure 1 is a typical V(I) characteristics in the absence of external noise obtained by dc measurements for sample #1. Above a threshold voltage $V_t$ = 9.58 mV corresponding to a threshold field $E_t$ = 95.8 mV/cm the transient response is initially fast followed by a slow increase, as observed previously by Fleming et al. [19].

The normalized conductivity $\sigma/\sigma_0$, with $\sigma_0$ the conductivity in the Ohmic regime, is plotted in Figure 2 (a) for samples #1, 2, 3, 4 and 5 as a function of the normalized voltage. Their threshold voltages are respectively : 9.58; 25, 19, 20 and 100 mV. The excess CDW conductivity is the highest in sample #5. Figure 2(b) shows $\sigma/\sigma_0$ forW-doped samples. The observed threshold voltage was 45 mV for x=0.001 and 114 mV for x=0.002 with corresponding threshold fields 450 and 1140 mV/cm, consistent with previous results [20]. The excess CDW conductivity is much smaller than in pure samples. For x=0.004, the nonlinear conductivity was highly metastable, as seen previously [21].

For the standard measurements of the BBN, a dc current was used. Figure 3 shows for the same sample #1 the rms voltage amplitude of the BBN in the frequency range 10 Hz–30 kHz as a function of dc current bias. Below threshold the BBN is negligible. A sharp onset of BBN appears at threshold, then the BBN shows a broad maximum.

For a given noise amplitude characterized by the modulation depth D, we have plotted in Figure 4 the ratio $\Delta V/V = [V_{pp}(D=0)-V_{pp}(D)]/V_{pp}(D=0)$ as a function of $V_{pp}(D=0)$ for sample #1 and for a noise modulation depth D=40%. $V_{pp}(D)$ is the response obtained after averaging over 2048 steps. $V_{pp}(D)$ is found to be always smaller than $V_{pp}(D=0)$. On sample #1, there is a non monotonous behavior of $\Delta V/V$ with $V_{pp}(D=0)$. $\Delta V/V$ starts to increase slightly below threshold then shows a broad maximum above $V_t$. There is clearly a residual background effect as seen in Figure 5 on the three Ohmic resistors r = 58, 25 and 6 Ohm for which no effect is expected. This Figure shows that the background effect does not depend on the value of the resistor.

Figures 6(a,b) illustrate the ratio $\Delta V/V$ as a function of $V_{pp}(D=0)$ for samples #2 with D=40% and D=80% and sample #3 with D=40% which show respectively threshold voltage $V_t = 25$ mV and $V_t = 19$ mV, corresponding to threshold fields 250 mV/cm and 190 mV/cm, higher than those of sample #1. $\Delta V/V$ increase near threshold and shows a broad maximum above.

Figure 7 shows $\Delta V/V$ as a function of $V_{pp}(D=0)$ for sample #4 with $V_t = 20$mV and $E_t = 200$ mV/cm and for different values of the modulation depth D=20, 40 and 60%. $\Delta V/V$ is nearly constant below threshold and increases noticeably above, corroborating results obtained on the other samples. Figures 6(a) and 7 show that the onset of the increase of $\Delta V/V$ does not depend on D. Figure 8 shows $\Delta V/V$ as a function of $V_{pp}(D=0)$ for sample #5 with a large threshold voltage $V_t =100$ mV. Figure 9 shows $\Delta V/V$ as a function of $V_{pp}(D=0)$ for W-doped samples. In both cases, no change is observed in $\Delta V/V$ near threshold.

Figure 10 shows $\Delta V/V$ vs D at two fixed values of $V_{pp}(D=0)/2$ below and above threshold for sample #4 and a resistor r= 58 Ohm. For $V_{pp}(D=0)/2 < V_t$, the curves for the resistor and sample #4 are very similar. For $V_{pp}(D=0)/2 > V_t$, the curve for sample #4 lies well above that obtained for $V_{pp}(D=0)/2 < V_t$ while the curve for the resistor remains very similar.

**Discussion**

The external applied white noise is a multiplicative noise since its amplitude D is proportional to the pulse amplitude. D represents a fluctuating force between V(1+D) and V(1-D) with $<D(t)> = 0$ and D(t) uncorrelated in time while the BBN is related to fluctuations of the barriers. The system is driven simultaneously by noise and a deterministic force V(t). The simplest interpretation for the increase of $\Delta V/V$ when the system is driven through the threshold field for depinning would be that the change of the chordal resistance at threshold due to the depinning of the CDW would lead to a change in $\Delta V/V$. This is not the case since, as it can be seen in Figure 5, the measurements of $\Delta V/V$ performed on the three Ohmic resistors r = 58, 25 and 6 Ohm show approximately the same dependence upon $V_{pp}(D=0)$. Since one does not expect change in $\Delta V/V$ on a resistor, this can be viewed as a residual background effect.

Also, as seen in Figure 10, $\Delta V/V$ measured as a function of D on the 58 Ohm resistor shows no dependence on $V_{pp}(D=0)$ while the $\Delta V/V$ vs D curves measured on sample #4 below and above threshold are noticeably different. The relative change of $\Delta V/V$ is more sensitive to D above threshold than below. The results indicate clearly an extra contribution due to the sliding CDW.

The excess CDW current in the presence of a noise $\delta V$ at a given applied voltage $V_0$ can be expressed as a Taylor expansion:

$f(V_0+\delta V) = f(V_0) + \delta V(df/dV) + \ldots + \delta V^n/n! \, (d^n f/dV^n) + \ldots$

and

$f(V_0-\delta V) = f(V_0) - \delta V(df/dV) + \ldots + (-1)^n \delta V^n/n! \, (d^n f/dV^n) + \ldots$

The average can be written:

½ $[f(V_0+\delta V)+f(V_0-\delta V)] = f(V_0) + \delta V^2/2 (d^2 f/dV^2) + \ldots + \delta V^{2p}/(2p!)(d^{2p} f/dV^{2p}) + \ldots$

Taking the expression for the CDW current $j_{CDW} \alpha (E/E_t-1)^\beta$ with $\beta \sim 1.4$ found by G. Mihaly et al. [17] it can easily be shown that all the terms $\delta V^{2p}/(2p!)(d^{2p} f/dV^{2p})$ are positive. Since the experimentally obtained average ½ $[f(V_0+\delta V)+f(V_0-\delta V)]$ is smaller than $f(V_0)$ the observed effect cannot be ascribed to a simple change in $d^{2p} f/dV^{2p}$ and seems to be due rather to an extra contribution of the sliding CDW induced by the presence of noise.

Moreover, if the change in ΔV/V was simply related to a change in the chordal resistance, one would expect a shift in the onset of the increase of ΔV/V upon varying D. This is not the case. Figures 6 (a,b) and 7 show that ΔV/V increases when $V_{pp}(D=0)/2$ is above threshold. This indicates that the excess CDW current increases when an external noise is added. We propose that noise is homogeneizing the CDW phase. This leads to a small increase of the nonlinear conductivity. This noise induced feature, namely a noise enhanced phase coherence, is reminiscent of stochastic resonance [22]. In nonlinear systems, this counterintuitive phenomenon arises from an interplay between noise and a deterministic external periodic signal, the repetitive pulse excitation in our experiments. Stochastic resonance can be viewed as an increase of the signal-to-noise ratio at the output through an increase of the noise level at the input of the nonlinear system. Here, noise cooperates with a coherent periodic excitation. In this context, the decrease of ΔV/V above ~$2V_t$ which appears in samples # 1 and # 3 (see Figures 4 and 6 (b)) and only above ~$4V_t$ in sample # 4 (see Figure 7) indicate that stochastic resonance effects would be observed in a larger interval in sample # 4 than in samples # 1 and # 3.

In the context of ratchets (the name for a periodic potential with no reflection symmetry) which show strong links with stochastic resonance [23], adding noise increases the rate of flow in a tilted ratchet in the overdamped regime up to a particular value. Further increase of noise leads to a decrease of the rate of flow [24]. In this picture, the decrease of ΔV/V above a sample dependent voltage value could be related to some changes in the asymmetry of the periodic pinning potential from sample to sample.

The ratio defined as :

$\Gamma = (\Delta V/V)_{max} / (\Delta V/V)_{V<V_t}$ and taken at D=40% from Figures 4, 6 (a,b), 7, 8 and 9 is shown in Figure 11 as a function of the threshold voltage $V_t$ of the corresponding samples. This ratio seems to be correlated to $V_t$, with the highest ratio obtained for sample #1 showing the lowest threshold voltage and being therefore of higher quality than the other samples. Samples #3 and #4 which show very similar threshold voltages but significantly different normalized conductivities $\sigma/\sigma_0$ also exhibit similar values of the ratio $\Gamma$. Even though sample # 5 shows a large normalized conductivity $\sigma/\sigma_0$, it exhibits a large threshold voltage and no change in ΔV/V. This corroborates the relationship between $V_t$ and $\Gamma$.

Numerical simulations of the effects of an additive noise on the CDW dynamics in the context of the Fukuyama-Lee-Rice model have been reported in ref. [25]. An increase of the spatial coherence of the phase of the CDW has been obtained. In other reports [26], a current noise of amplitude proportional to the CDW current has been used in a Langevin equation for the overdamped CDW phase dynamics and in the associated Fokker-Planck equation.

The absence of change of $\Delta V/V$ near threshold in W-doped samples might be rather related to the very weak CDW conductivity, as seen in Figure 2(b).

One should note that, in sample # 1, $\Delta V/V$ seems to increase slightly below threshold. This may reveal excitation of metastable CDW configurations in the pinned state induced by the external noise.

**Conclusion**

In summary, we have experimentally investigated for the first time nonlinear effects of a multiplicative external noise on the CDW driven by a rectangular bipolar signal in pure and W-doped $K_{0.30}MoO_3$. We have proposed that the non monotonous behaviour of $\Delta V/V$ with V(D=0) for a given noise modulation depth D can be described qualitatively in terms of stochastic resonance. We have found that $\Delta V/V$ is correlated to the threshold field $E_t$ for depinning with a decrease of $\Delta V/V$ when $E_t$ increases. A precise mechanism for the relative change in the amplitude of the voltage pulse $\Delta V/V$ as a function of noise amplitude and bias remains to be clarified. Theoretical studies of the effect of adding a white multiplicative noise to the transient CDW response would be necessary.

**Figure 1.** V(I) characteristics measured with dc current at T=77K for sample #1.

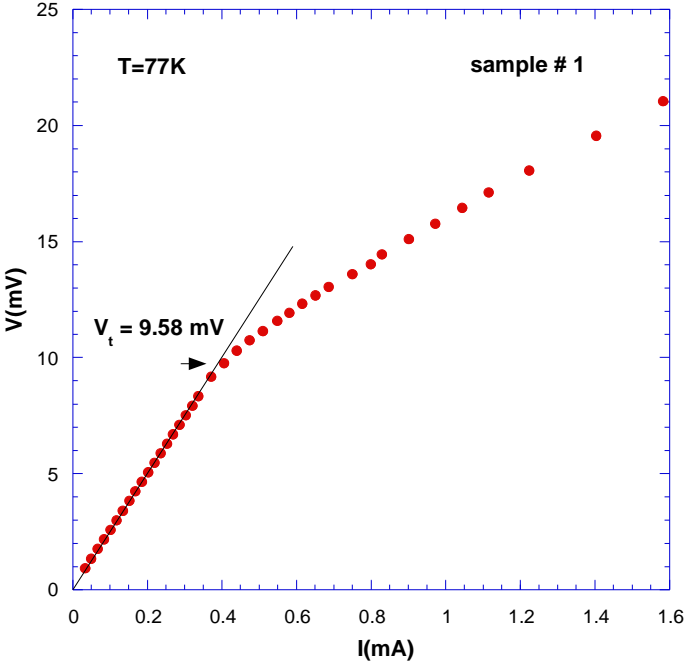

**Figure 2(a)**. Normalized conductivity σ/σ₀ as a function of normalized threshold voltage V/V$_t$ for samples #1, #2, #3, #4 and #5. Threshold voltages: V$_t$ = 9.58 mV (sample #1), 25 mV (sample #2), 19 mV (sample# 3), 20 mV (sample #4), 100 mV (sample #5).

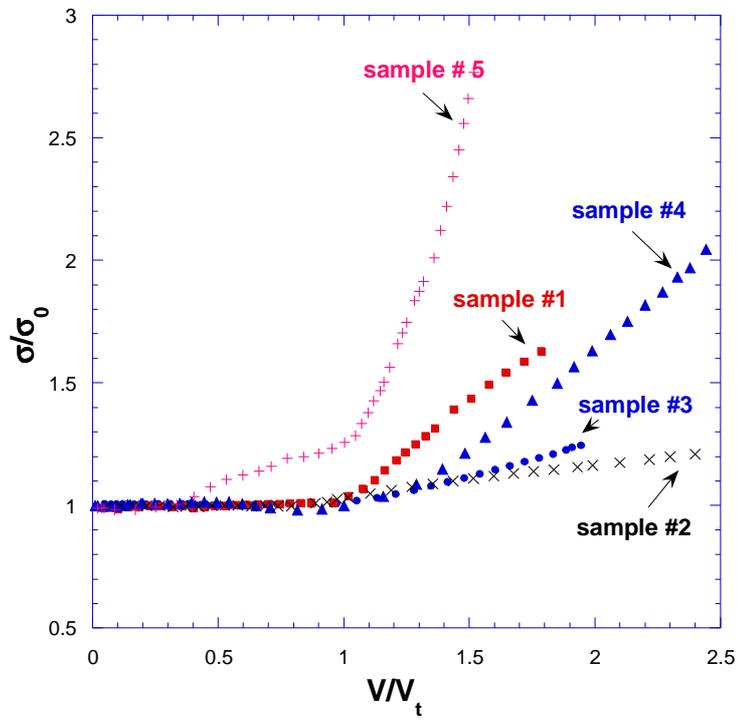

**Figure 2(b).** Same as in Figure 2(a) for W-doped samples ; x=0.001 ($V_t$=45 mV) and x=0.002 ($V_t$=114 mV).

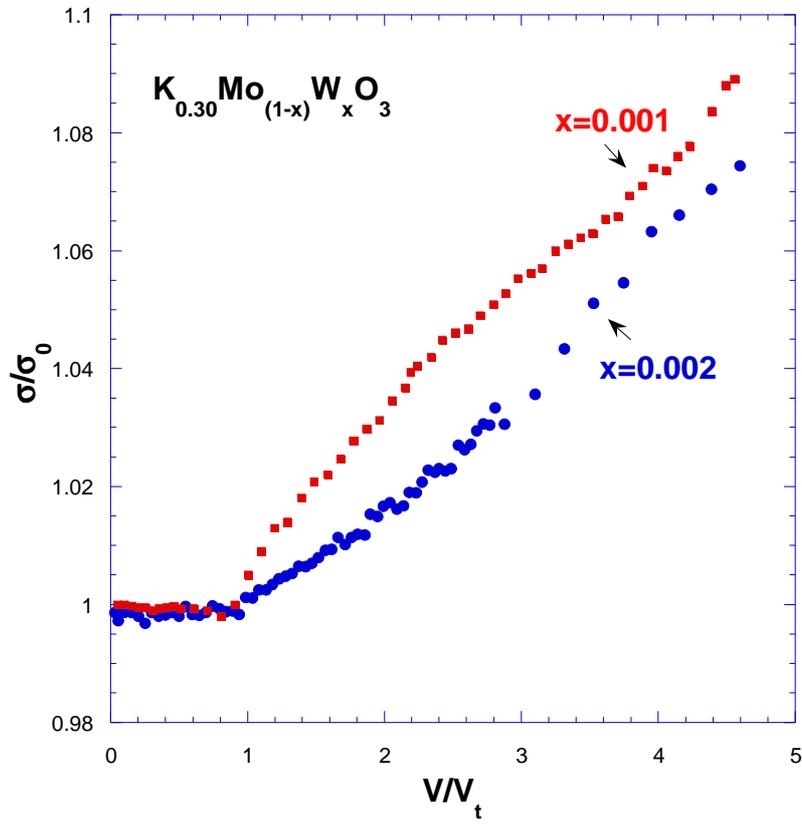

**Figure 3**. rms broad band noise voltage amplitude as a function of dc current bias in the frequency range 10Hz-30kHz for sample # 1.

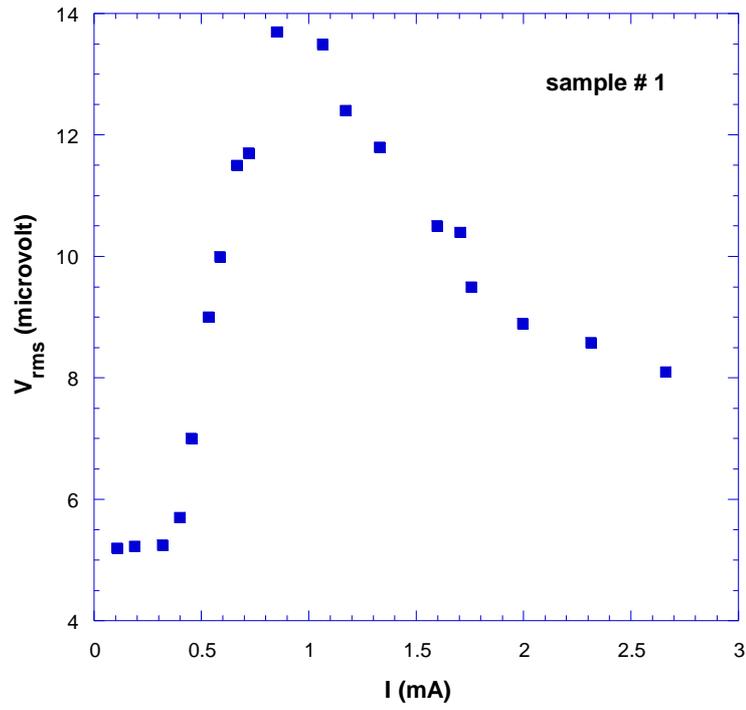

**Figure 4.** [$V_{pp}(D=0)-V_{pp}(D)$] /$V_{pp}(D=0)$ as a function of $V_{pp}(D=0)$ for noise modulation depth $D = 40\%$ for sample # 1. Insert: sketch of the rectangular periodic bipolar signal.

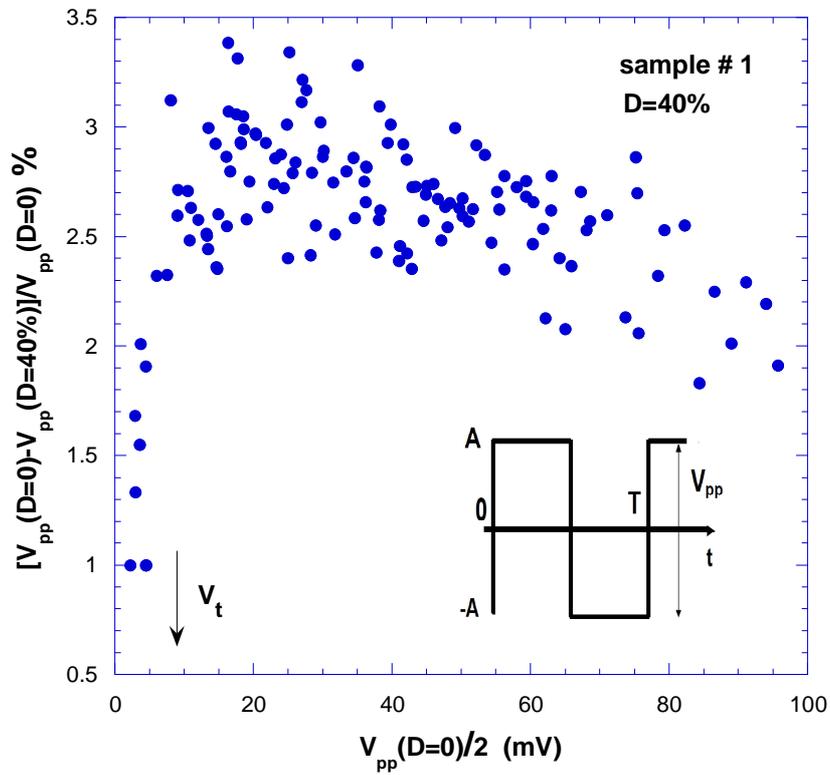

**Figure 5.** [$V_{pp}(D=0)-V_{pp}(D)$] /$V_{pp}(D=0)$ as a function of $V_{pp}(D=0)$ for three resistors : 58, 25 and 6 Ohm.

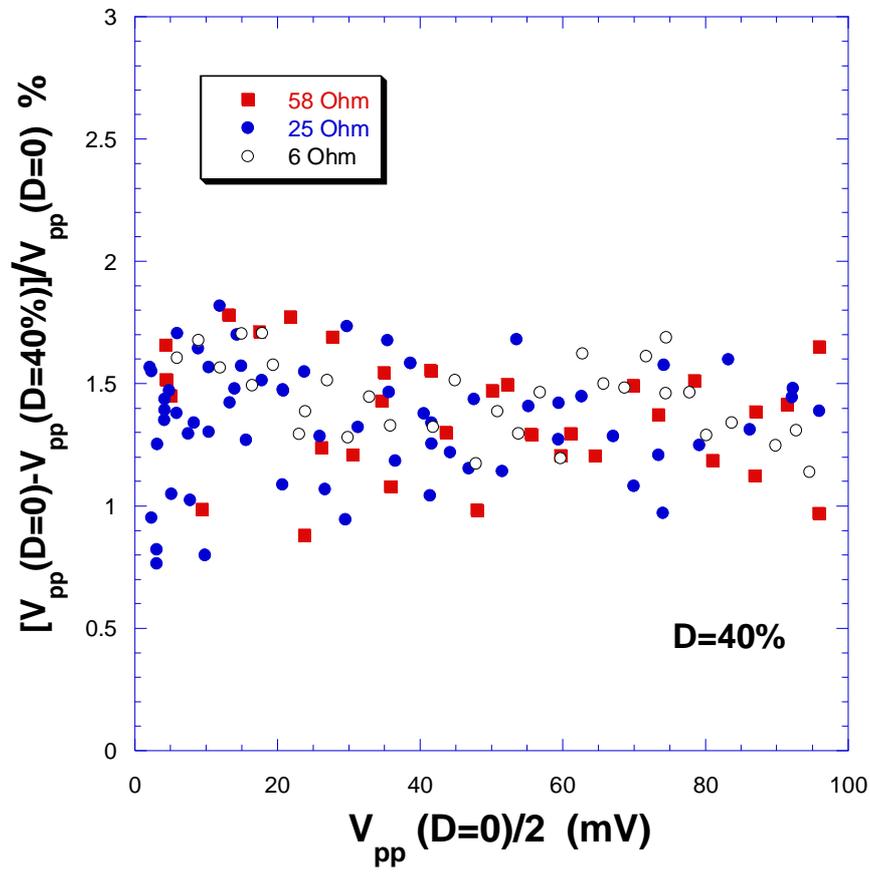

**Figure 6(a).** $[[V_{pp}(D=0)-V_{pp}(D)]/V_{pp}(D=0)$ as a function of $V_{pp}(D=0)$ for D=40% and D=80% for sample # 2. $V_t$ =25 mV.

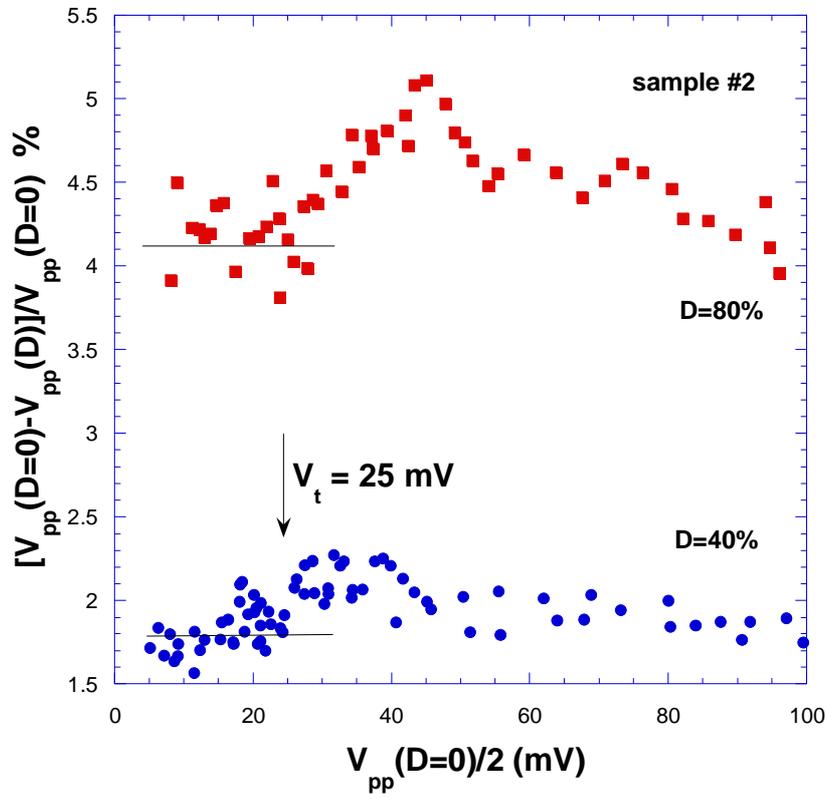

**Figure 6(b).** Same as in Figure 6(a) for sample # 3 for D=40%. $V_t$ =19 mV.

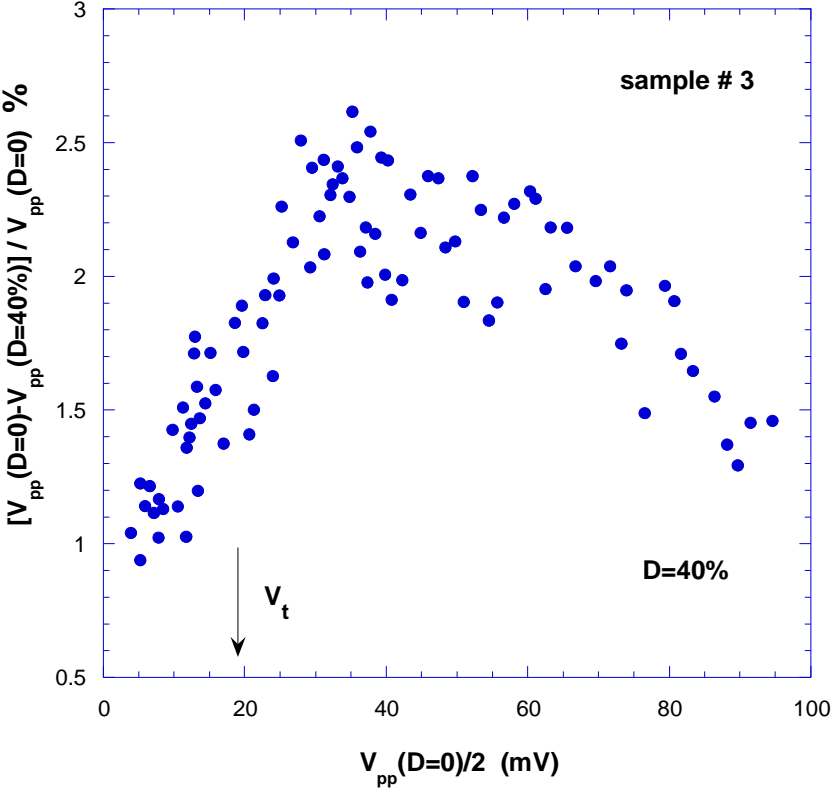

**Figure 7.** [$V_{pp}(D=0)-V_{pp}(D)$] /$V_{pp}(D=0)$ as a function of $V_{pp}(D=0)$ for D = 20, 40, 60% for sample # 4. Straight lines are guides to the eye.

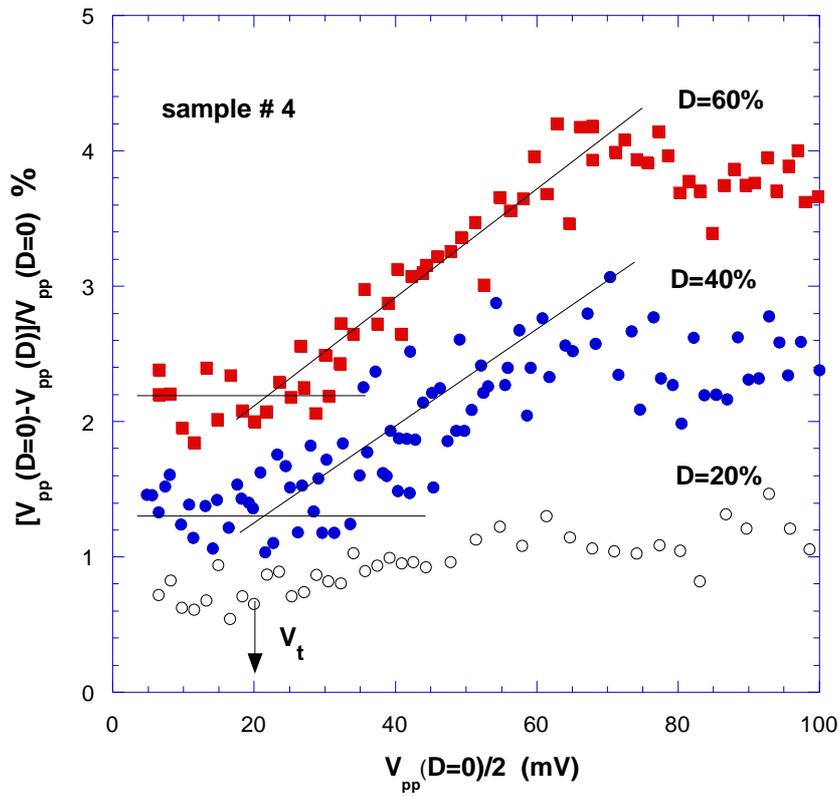

**Figure 8**. Same as in Figure 6(a) for sample # 5 for D=40%. $V_t$ =100 mV.

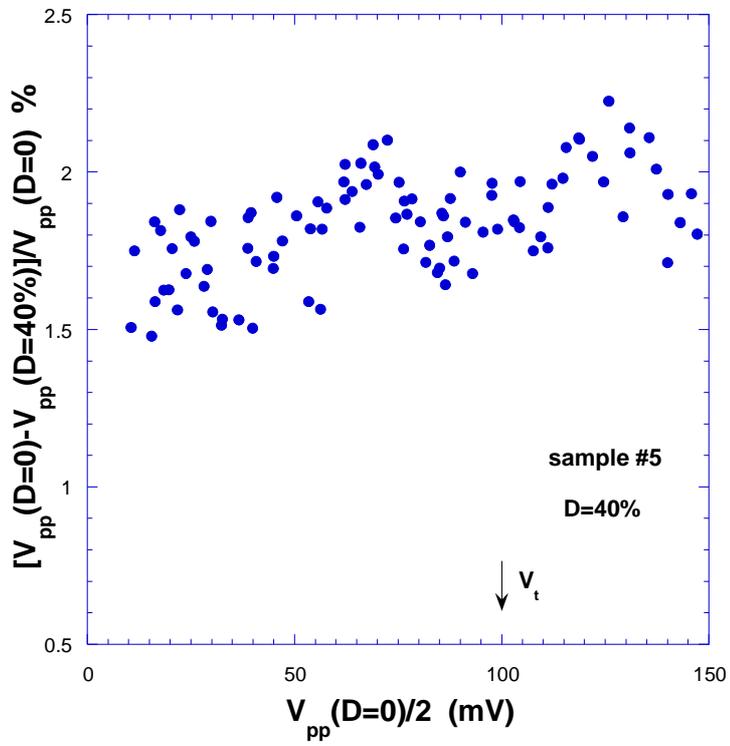

**Figure 9.** [$V_{pp}(D=0)-V_{pp}(D)$] /$V_{pp}(D=0)$ as a function of $V_{pp}(D=0)$ for D = 40% for W-doped samples, x=0.001 and x=0.002.

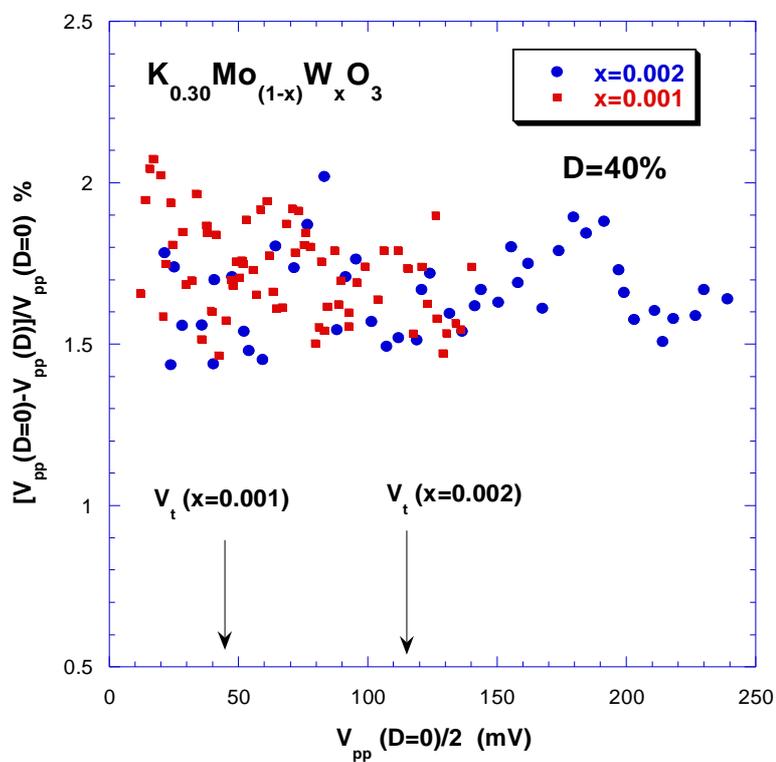

**Figure 10.** [$V_{pp}$(D=0)-$V_{pp}$(D)] /$V_{pp}$(D=0) as a function of D for resistor r=58 Ohm and for sample # 4 for two values of $V_{pp}$(D=0)/2 : 4.78 mV, below threshold and 67.85 mV, above threshold for sample # 4.

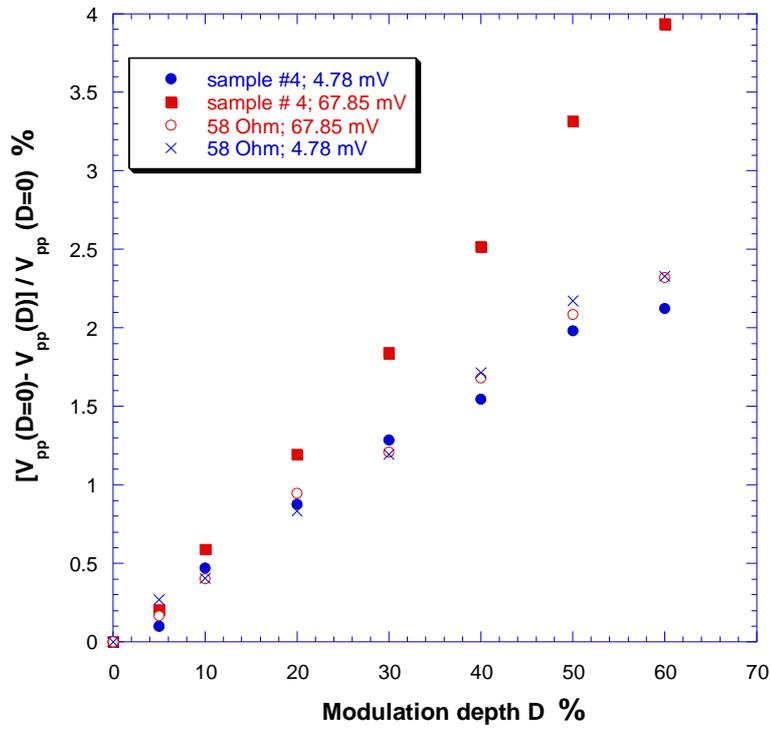

**Figure 11**. $(\Delta V/V)_{max}/\Delta V/V)_{V<V_t}$ as a function of $V_t$ for different samples. D=40%. See text for discussion.

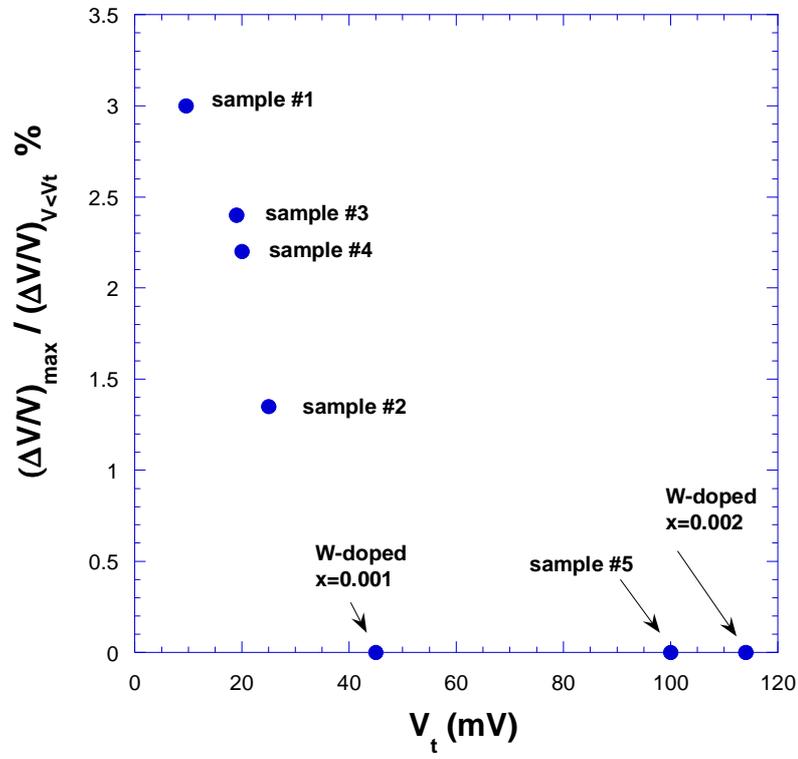